\documentclass[10pt]{article}

\begin{document}

\title{Mass and Charge in Brane-World and Non-Compact Kaluza-Klein Theories in 5 Dim}
\author{J. Ponce de Leon\thanks{E-mail: jponce@upracd.upr.clu.edu}\\ Laboratory of Theoretical Physics, Department of Physics\\ 
University of Puerto Rico, P.O. Box 23343, San Juan, \\ PR 00931, USA }
\date{March 2003}

\maketitle
\begin{abstract}
In classical Kaluza-Klein theory, with compactified extra dimensions and without scalar field, the rest mass as well as the electric charge of test particles are constants of motion. We show that in the case of a large extra dimension this is no longer so.
We propose the Hamilton-Jacobi formalism, instead of the geodesic equation, for the study of test particles moving in a five-dimensional background metric. This formalism has a number of advantages: (i) it provides a clear and invariant definition of rest mass, without the ambiguities  associated with the choice of the parameters used along the motion in $5D$ and $4D$, (ii) the electromagnetic field can be easily incorporated in the discussion, and (iii) we avoid the difficulties associated with the  ``splitting" of the geodesic equation. For particles moving in a general $5D$ metric, we  show how the effective rest mass, as measured by an observer in $4D$, varies as a consequence of the large extra dimension. Also, the fifth component of the momentum changes along the motion. This component can be identified with the electric charge of test particles.  With this interpretation, both the rest mass and the charge vary along the trajectory.  The constant of motion is now a combination  of these quantities. We study the cosmological variations of charge and rest mass in a five-dimensional bulk metric which is used to embed the standard $k = 0$ FRW universes. The time variations in the fine structure ``constant" and the Thomson cross section are also discussed. 

 \end{abstract}



\newpage

\section{Introduction}
Because of the cylindricity condition, in classical or {\em compactified} Kaluza-Klein theory, the ``new" physics predicted by the five-dimensional equations, resides completely in the four potentials of electromagnetism $A_{\mu}$ and the scalar potential $\Phi$. The $4D$ part of the five-dimensional equations can be manipulated to reproduce the four-dimensional physics with some new terms, or corrections. If the electromagnetic potentials are absent and the scalar field is constant, then these new terms  vanish identically. 

However, the cylindricity condition is not required, nor, in general, sustained. The possibility  that our world might contain more than four {\em non-compact} dimensions has attracted the interest of a great number of researchers. 

In higher-dimensional gravity theories, inspired by string theories, 
instead of cylindricity the matter fields are assumed to be confined to our $4D$ spacetime, which is modeled as a singular hypersurface or ``brane" embedded in a larger $(4 + d)$ world. In these theories gravity is a multidimensional interaction that can propagate in the extra $d$ dimensions as well \cite{Arkani1}-\cite{Arkani3}. In the RS scenario, for $d= 1$, the usual general relativity in $4D$ can be recovered even when the extra dimension is infinite in size \cite{RS1}-\cite{RS2}. 

In space-time-matter theory (STM), inspired by the unification of matter and geometry, instead of cylindricity the matter fields in our $4D$ spacetime are assumed to be ``induced" or derived from pure geometry in five dimensions \cite{Wesson 1}-\cite{Wesson book}. Our $4D$ spacetime can, in principle, be {\em any} four-dimensional hypersurface orthogonal to the extra dimension, not necessarily a singular one \cite{JPdeL 1}-\cite{JPdeL Wesson}.

The existence of a large extra dimension is intriguing and provides a wealth of new physics, even in the absence  of electromagnetic field and constant scalar potential. This is illustrated by  the effective equations for gravity in $4D$, which predict five-dimensional local and non-local corrections to the usual general relativity in $4D$ \cite{Shiromizu}-\cite{Deruelle and Katz}. Also, the geodesic equation for test particles in $5D$ predicts an effective four-dimensional equation of motion with an extra non-gravitational force \cite{MashhoonWesson}-\cite{Seahra3}. This force, called sometimes ``fifth" force,  has a component that is a direct consequence of the variation of the rest mass of test particles, induced by the large extra dimension. 

In this work we study in more detail the effects of a large extra dimension on the rest mass and electric charge of test particles.  The possibility that these quantities  might be variable has important implications for the foundations of physics because variable mass and/or charge imply time-varying Thomson cross section  $\sigma = (8\pi/3)(q_{e}^2/m_{0}c^2)^2$ for the scattering of electromagnetic radiation by a particle of charge $q_{e}$ and mass $m_{0}$. Also the variation of electric charge $q_{e}$ implies the variation of the electromagnetic fine structure ``constant" $ \alpha_{em} = q_{e}^2/(4 \pi \hbar c)$. The latter has attracted considerable attention in view of the recent observational evidence that $\alpha_{em}$  might vary over cosmological time scales \cite{Webb1}-\cite{Webb2}. This, of course,  requires the time variation of at least one of the ``constants" ($q_{e}$, $\hbar$ and $c$). However, recently a number of theories attribute the variation of the fine structure constant to changes in the fundamental electron charge and preserve $c$ (Lorentz invariance) and $\hbar$ as constants \cite{Barrow1}-\cite{Youm4}.

We have other theoretical motivations for this study. Namely, that in compactified Kaluza-Klein theory, with the imposition of the so-called cylinder condition, the electric charge is strictly constant. Therefore, we would like to see whether a large extra dimension would once more  give new physics and change the classical scenario.

Besides, we have some technical motivations. Firstly, previous results regarding the variation of rest mass of test particles were  obtained from the analysis of the geodesic equation \cite{MashhoonWesson}-\cite{Seahra3}, \cite{Seahra4}-\cite{Billyard}. However, the mass of the test particle appears nowhere  in this equation. 
Secondly, the interpretation of the rest mass in terms of the extra coordinate, based on the geodesic equation, seems to depend heavily on the choice of system of coordinates and the choice of parameter used to characterize the motion \cite{Youm1}, \cite{Seahra4}, \cite{Seahra2}.

What we propose here is to define the properties  
of test particles, as measured by an observer in $4D$, through the Hamilton-Jacobi equation, instead of the geodesic one. 

This approach has several advantages. From a physical viewpoint, it provides a clear and invariant definition of rest mass, without the problems associated with the parameter used along the motion. In addition the electromagnetic field can be easily incorporated in the discussion. From a mathematical viewpoint, this is due  to the fact that the Hamilton-Jacobi equation is a  {\em scalar} equation. We thus avoid the complications associated with the  $(4 + 1)$ ``splitting " of the $5D$ geodesic equation\footnote{It might be worth mentioning that, the  $(4 + 1)$ splitting of the five-dimensional equations with non vanishing electromagnetic potentials,  is definitely more involved  in the case of a large extra dimension than in the classical case with compact extra dimensions \cite{Eq. of Motion}, \cite{EMT}.}. 

In this work we obtain the general expression for the change of the rest mass of a test particle, as perceived by an observer in $4D$. It includes the combined effects from a large extra dimension, and those from the  electromagnetic and scalar potentials. 

We also show that the fifth component of the momentum, which in $4D$ can be identified with the electric charge of the test particle,  changes along the motion. This effect has no counterpart in classical  theory, where the charge is strictly constant. Thus, a large extra dimension leads naturally to a variation of the fine structure constant. For the case where $c$ and $\hbar$ are taken to be constants, we present the equation for the relative variation of $\alpha_{em}$.

The conservation equation for mass and charge, as well as the relationship in their variation, is also discussed.  

\section{The Formalism}

In this section we give a brief review  of the Kaluza-Klein equations in terms of local basis vectors, which are used to project five-dimensional quantities onto spacetime. The projection of the five-momentum on $4D$ provides the correct definition for the  rest mass. The Hamilton-Jacobi equation in $4D$ suggests we interpret the extra component of the five-momentum as the electric charge of the test particle. Finally, a  conservation equation for the mass and charge follows from the five-dimensional Hamilton-Jacobi.

\subsection{Line element in Kaluza-Klein theory}
We will consider a general five-dimensional manifold with coordinates $\xi^{A}$ $(A = 0,1,2,3,4)$ and metric tensor $\gamma_{AB}(\xi^{C})$. The $5D$ interval is then given by
\begin{equation}
\label{general 5D metric}
d{\cal S}^2 = \gamma_{AB}d{\xi}^A d{\xi}^B.
\end{equation}
We assume that the $5D$ manifold in (\ref{general 5D metric}) allows us to construct, a four-dimensional hypersurface that can be identified with our $4D$ space-time. In this hypersurface we can introduce an arbitrary  set of coordinates  $x^{\mu}$ $(\mu= 0,1,2,3)$, which are functions of $\xi^A$,
\begin{equation}
\label{4D coordinates}
x^{\mu}= x^{\mu}\left({\xi^0},{\xi^1},{\xi^2},{\xi^3},{\xi^4}\right).
\end{equation}
The simplest choice is the ``coordinate frame", where the first four coordinates $\xi^{\mu}$ are assumed to be the coordinates of spacetime

\begin{equation}
\label{special coordinates}
x^{\mu}= {\xi}^{\mu}.
\end{equation}
The corresponding spacetime basis vectors, $ {\hat{e}}^{(\mu)}_{A} = \partial x^{\mu}/\partial \xi^A$,  are
\begin{eqnarray}
\label{space-time basis vectors}
\hat{e}^{(0)}_{A}&=& (1, 0, 0, 0, 0),\nonumber \\
\hat{e}^{(1)}_{A}&=& (0, 1, 0, 0, 0),\nonumber \\
\hat{e}^{(2)}_{A}&=& (0, 0, 1, 0, 0),\nonumber \\
\hat{e}^{(3)}_{A}&=& (0, 0, 0, 1, 0).
\end{eqnarray}
The vector $\psi^A$, orthogonal to spacetime is given by
\begin{equation}
\label{e4}
{\psi^A}= (0, 0, 0, 0, {\Phi}^{- 1}),
\end{equation}
where we have set $\gamma_{44}= \epsilon \Phi^{2}$, so that $\gamma_{AB}\psi^A\psi^B = \epsilon$. The factor $\epsilon$ is taken to be $+ 1$ or $- 1$ depending on whether the extra dimension is timelike or spacelike, respectively. Denoting $\gamma_{\mu 4}= \epsilon \Phi^2 A_{\mu}$, the associated basis vectors ${\hat{e}}^{A}_{(\mu)}$ are given by,
\begin{eqnarray}
\label{associated basis vectors}
\hat{e}^{A}_{(0)}&=& (1, 0, 0, 0, -A_{0}),\nonumber \\ 
\hat{e}^{A}_{(1)}&=& (0, 1, 0, 0, -A_{1}),\nonumber \\ 
\hat{e}^{A}_{(2)}&=& (0, 0, 1, 0, -A_{2}),\nonumber \\ 
\hat{e}^{A}_{(3)}&=& (0, 0, 0, 1, -A_{3}). 
\end{eqnarray}
They satisfy $\hat{e}^{(\mu)}_{A}\hat{e}^{A}_{(\nu)} = \delta^{\mu}_{\nu}$ and $\psi_{A}\hat{e}^{A}_{\mu} = 0$.  Also
\begin{equation}
\psi_{A}= \epsilon \Phi(A_{0}, A_{1}, A_{2}, A_{3}, 1).
\end{equation}
With this choice, spacetime displacements are $dx^{\mu} = \hat{e}^{(\mu)}_{A} d\xi^A = d\xi^{\mu}$, while displacements orthogonal to spacetime, which we will denote as $dx^4$,  are given by
\begin{equation}
dx^{4}= \psi_{A}d{\xi^A}= \epsilon \Phi(d{\xi}^4 + A_{\mu}dx^{\mu}).
\end{equation}
Since $d\xi^A = \hat{e}^{A}_{(\mu)}dx^{\mu} + \epsilon \psi^A dx^4$, the notation becomes more symmetrical in terms of the vector $\hat{e}_{(4)}^A = \epsilon \psi^A$, $\hat{e}^{(4)}_{A} = \psi_{A}$. Thus, $dx^M = \hat{e}^{(M)}_{A}d\xi^A$, $d\xi^B = \hat{e}^{B}_{(C)}dx^C$ and the five-dimensional metric, in the local frame, becomes
\begin{equation}
\label{gAB}
\hat{g}_{AB} = \hat{e}_{(A)}^C\hat{e}_{(B)}^D \gamma_{CD}.
\end{equation}
Conversely, $\gamma_{AB} = \hat{e}_{A}^{(M)}\hat{e}^{(N)}_{B}\hat{g}_{MN}$.
Therefore, the line element (\ref{general 5D metric}) takes the well known form  
\begin{eqnarray}
\label{special metric} 
d{\cal S}^2 &=& g_{\mu\nu}dx^{\mu}dx^{\nu}+ \epsilon \Phi^2\left(d{\xi}^4 + A_{\mu}dx^{\mu}\right)^2 = ds^2 + \epsilon (dx^{4})^2,\nonumber \\
g_{\mu\nu}&=& \gamma_{\mu\nu}-\epsilon\Phi^2A_{\mu}A_{\nu},
\end{eqnarray}
where  $ds$ is the four-dimensional interval. We note that the interpretation of the four-vector $A_{\mu}$ strongly relies on the signature of the extra dimension. If it is spacelike, ($\epsilon = -1$), then $A_{\mu}$ can be interpreted as the usual  four-potential of electromagnetism. If it is timelike, ($\epsilon = +1$), this interpretation fails  because we would have the wrong sign for the energy-momentum tensor of electromagnetism.     

\subsection{The Hamilton-Jacobi equation}
\subsubsection{General case}

The momentum $P^A$ of a test particle in a five-dimensional world is defined in the usual way, namely,
\begin{equation}
\label{5D Momentum}
P^{A} = M_{(5)} U^A, 
\end{equation}
where $U^A = d\xi^A/d{\cal{S}}$ is the $5D$-velocity, and $M_{(5)}$ is the $5D$ ``mass" of the particle. Thus, 
\begin{equation}
\gamma_{AB}P^AP^B = M_{(5)}^2 
\end{equation}
 If $S$ denotes the action as a function of coordinates, then substituting $ - \partial S/ \partial \xi^A$ for $P_{A}$, we obtain the Hamilton-Jacobi equation for a test 
particle in $5D$ 
\begin{equation}
\label{HJ}
\gamma^{AB}\left(\frac{\partial S}{\partial \xi^A}\right)\left(\frac{\partial S}{\partial \xi^B}\right) = M^2_{(5)},
\end{equation}

The four-dimensional momentum $P_{(\mu)}$ is the projection of the five-dimensional quantity $P_{A}$  onto the $4D$ spacetime, viz.,
\begin{equation}
\label{four momentum}
P_{(\mu)} =  {\hat{e}}_{(\mu)}^A P_{A} = - {\hat{e}}_{(\mu)}^A\frac{\partial S}{\partial \xi^A}.
\end{equation}
With this definition, $P_{(\mu)}$ is invariant under arbitrary transformations of coordinates ${\xi}^A = {\xi}^{A}({\bar{\xi}}^C)$ in $5D$, and behaves as a four-dimensional vector under $4D$ transformations $x^{\mu} = x^{\mu}({\bar{x}}^{\lambda})$. In addition, $m_{0}$, the effective rest mass  measured by an observer in $4D$, is invariantly defined as  
\begin{equation}
\label{def of rest mass}
g^{\mu\nu}P_{(\mu)}P_{(\nu)} = m_{0}^2.
\end{equation}
This definition is entirely free of the problems mentioned in the introduction. Namely, the rest mass (\ref{def of rest mass}) is independent on the particular set of coordinates and  the choice of parameter used to characterize the motion.

  From (\ref{gAB}), it follows that $\gamma^{AB} = {\hat{e}}^{A}_{(M)}{\hat{e}}^{B}_{(N)}\hat{g}^{MN}$. Using this in (\ref{HJ}) we obtain 
\begin{equation}
\label{HJ in orth. frame}
g^{\mu\nu}\left(\hat{e}_{(\mu)}^A \frac{\partial S}{\partial \xi^A}\right)\left({\hat{e}}_{(\nu)}^{B}\frac{\partial S}{\partial \xi^B}\right) + \epsilon {\hat{e}}_{(4)}^A {\hat{e}}_{(4)}^B \frac{\partial S}{\partial \xi^A}\frac{\partial S}{\partial \xi^B} = M_{(5)}^2
\end{equation}
Now the substitution of (\ref{four momentum}) and (\ref{def of rest mass}) into (\ref{HJ in orth. frame}) yields
\begin{equation}
\label{rel for m, M and q}
m_{0}^2 + \epsilon\frac{q^2}{\Phi^2} = M_{(5)}^2, 
\end{equation}
where we have used the notation
\begin{equation}
\label{def of q}
q = P_{4} = - \frac{\partial S}{\partial y}.
\end{equation}
Here, and in the remainder of this note, we use $y$ instead of $\xi^{4}$, namely $y = \xi^4$.
From (\ref{space-time basis vectors}), (\ref{5D Momentum}) and (\ref{def of q}) it follows that
\begin{equation}
\label{q in terms of u4}
q = \frac{M_{(5)} \Phi u^4}{\sqrt{1 + \epsilon (u^{4})^2}},
\end{equation}
where 
\begin{equation}
\label{u4}
u^{4} = \frac{dx^{4}}{ds} = \epsilon \Phi \left(\frac{dy}{ds} + A_{\mu} u^{\mu}\right).
\end{equation}
 In order to avoid misunderstanding, we stress the fact that $u^4$ is not a part of the four-velocity vector, which is $u^{\mu} = dx^{\mu}/ ds$. Since the displacement orthogonal to spacetime is given by $dx^{4} = {\hat{e}}^{(4)}_{A}d \xi^A$ ({\em not} by $dy = d \xi^4$), it follows that $u^4$ characterizes the velocity,  of a test particle moving in the bulk metric, orthogonal to spacetime.

A simple expression relating $m_{0}$, the effective rest mass in $4D$, and $M_{(5)}$ can be obtained by substituting (\ref{q in terms of u4}) into (\ref{rel for m, M and q}), namely, 
\begin{equation}
\label{m in terms of u4}
m_{0} = \frac{M_{(5)}}{\sqrt{1 + \epsilon (u^{4})^2}}.
\end{equation}
It shows that the rest mass of a particle, perceived by an observer in $4D$, varies as a result of the five-dimensional motion along the extra direction. This is similar to $m = m_{0}/ \sqrt{1 - v^2}$, for the change of mass  due the motion in space. From (\ref{q in terms of u4}) and (\ref{m in terms of u4}) we get
\begin{equation}
\label{q in terms of m and Phi}
q = m_{0} \Phi u^{4}.
\end{equation}
In the case where $M_{(5)} \neq 0$, equations (\ref{rel for m, M and q}) and (\ref{q in terms of m and Phi}) constitute a set of two linearly independent equations. They form  the basis for our discussion in the next section. 

\subsubsection{Massless particles in $5D$}

The motion of massless particles is along isotropic geodesics. Since for such geodesics  $d{\cal{S}} = 0$, from (\ref{special metric}) it follows that
\begin{equation}
\label{ds for null geodesics}
ds^2 = - \epsilon (dx^4)^2.
\end{equation}
Particles with real mass (instead of imaginary mass) follow timelike geodesics, for which $ds^2 > 0$. Thus, in the case of a spacelike extra dimension, there are two physical possibilities: 

(i) If the particle has non-vanishing motion perpendicular to spacetime, then a massless particle moving in $5D$ is perceived  as a massive particle by an observer in $4D$.

(ii) If the motion in $5D$ is longitudinal to spacetime ($dx^4 = 0$), then $ds = 0$ indicating that a massless particle in $5D$, is observed as massless particle in $4D$. 

In the case of a timelike extra dimension, from  $m_{0}^2 + q^2/\Phi^2 = 0$ it follows that there is only one physical possibility, viz.,  $m_{0} = q = 0$.

The equation of motion for massless particles is the Eikonal equation, which differs form the one of Hamilton-Jacobi in that, in the right hand side of (\ref{HJ}) we set $M_{(5)} = 0$. Also, in (\ref{5D Momentum})  the derivatives $M_{(5)}d/d{\cal{S}}$  ought to be replaced by $d/d\lambda$, where $\lambda$ is the parameter along the null geodesic \cite{Landau and Lifshitz}. The effective rest mass in $4D$ is still defined by (\ref{def of rest mass}), and (\ref{rel for m, M and q}) yields ($\epsilon = -1$)
\begin{equation}
\label{massless 5D particles}
q = \pm m_{0}\Phi,
\end{equation}
which is the same as in (\ref{q in terms of m and Phi}), if we notice from (\ref{ds for null geodesics}) that now $u^4 = \pm 1$.      

\subsection{Interpretation of $q$}

For the physical interpretation of the fifth component of $P_{A}$, or shorter of $q$, we note that in the coordinate frame (\ref{associated basis vectors}), the four-momentum (\ref{four momentum}) becomes
\begin{equation}
\label{gen. Momentum} 
P_{(\mu)} = - \frac{\partial S}{\partial x^{\mu}} + A_{\mu} \frac{\partial S}{\partial y}.
\end{equation}
Then, the explicit form of (\ref{def of rest mass}), in the coordinate frame, is as follows
\begin{equation}
g^{\mu\nu}\left(\frac{\partial S}{\partial x^{\mu}} - A_{\mu}\frac{\partial S}{\partial y}\right) \left(\frac{\partial S}{\partial x^{\nu}} - A_{\nu}\frac{\partial S}{\partial y}\right) = m_{0}^2.
\end{equation}
For a spacelike extra dimension, $A_{\mu}$ can be interpreted as the electromagnetic potential. On the other hand, the Hamilton-Jacobi equation for a particle with electric charge $q_{e}$ moving in an electromagnetic field is 
\begin{equation}
g^{\mu\nu}\left(\frac{\partial S}{\partial x^{\mu}} + q_{e} A_{\mu}\right) \left(\frac{\partial S}{\partial x^{\nu}} + q_{e} A_{\nu}\right) = m_{0}^2.
\end{equation}
The comparison of the last two equations suggests, that for $\epsilon = -1$, we identify $ P_{4} = - \partial S/\partial y$ with the electric charge of the test particle, viz., 
\begin{equation}
\label{def of charge}
q = q_{e}.
\end{equation}
Thus, the motion of a test particle in the background metric, along the {\em spacelike} extra dimension, is perceived by an observer in $4D$ as the electric charge. 
This interpretation is possible because, $q = P_{4}$ and $q_{e}$ remain invariant in all frames of reference in $4D$, i.e., they are scalars under arbitrary coordinate transformations in spacetime. 

From (\ref{q in terms of m and Phi}), or from (\ref{massless 5D particles}) for massless $5D$ particles, it follows that the electric charge $q_{e}$ is proportional to the scalar field $\Phi$ which varies in space and time. Note that the charges of all particle species vary in the same way (so that, for example, atoms can remain always neutral \cite{Youm3}-\cite{Youm4}). This is similar to the assumption made in cosmologies with varying $q_{e}$ \cite{Barrow1}-\cite{Youm4}, \cite{Bekenstein}.  

In the case of a timelike extra dimension the interpretation of $P_{4}$ in terms of four-dimensional physics is not so clear. However,  there are exact solutions of the field equations of  $5D$ relativity,  with good physical properties and  a timelike extra dimension, for which the above interpretation seems to be applicable, with no contradictions \cite{int.sol}. 

\section{Effects of Large Extra Dimensions}

In this section we find the  general equations for the change of $m_{0}$ and $q$, in the presence of an electromagnetic field and scalar potential. We present the expression for the relative variation of the fine structure constant $\alpha_{em}$, corresponding to the case where $c$ and $\hbar$ are taken to be constants. Next, we consider the $5D$ metric in gaussian normal coordinates, which allows us  to isolate the effects produced by a large  extra coordinate, from those induced by the electromagnetic and scalar fields. Finally, we discuss the relationship between the changes of $m_{0}$ and $q$.  

\subsection{Variation of $m_{0}$ and $q$}

In order to find out whether the values of  $q$ and $m_{0}$ of a test particle are affected by a large extra dimension, we should find $dq/ds$ and $dm_{0}/ds$. 

First, we consider the motion of a massive $5D$-test particle moving in an arbitrary five-dimensional background metric. If $M_{(5)} \neq 0$, then  the equations (\ref{rel for m, M and q}) and (\ref{q in terms of m and Phi}) are linearly independent. Thus taking derivatives we find  

\begin{equation}
\frac{dq}{ds} = m_{0}\Phi \left[ \frac{1}{[1 + \epsilon (u^{4})^2]} \frac{du^{4}}{ds} + \frac{u^{4}}{\Phi} \frac{d\Phi}{ds}\right],
\end{equation}
\begin{equation}
\frac{dm_{0}}{ds} = - \frac{\epsilon q}{\Phi [1 + \epsilon (u^{4})^2]} \frac{du^{4}}{ds},
\end{equation}
We now need the expression for $du^{4}/ds$. It is given by equation (82) in Ref. \cite{Eq. of Motion}, namely
\begin{equation}
\frac{1}{[1 + \epsilon (u^{4})^2]}\frac{du^{4}}{ds} = \frac{1}{2\Phi}\left(\frac{\partial g_{\mu\nu}}{\partial y}\right)u^{\mu}u^{\nu} + \left( \frac{\partial A_{\mu}}{\partial y} + \frac{A_{\mu}}{\Phi} \frac{\partial \Phi}{\partial y} - \frac{1}{\Phi}\frac{\partial \Phi}{\partial x^{\mu}}\right)u^{\mu}u^{4},
\end{equation}
Substituting we get
\begin{equation}
\label{q dot}
\frac{dq}{ds} = \frac{1}{2}m_{0}u^{\mu}u^{\nu}\frac{\partial g_{\mu\nu}}{\partial y} + q u^{\mu}\frac{\partial A_{\mu}}{\partial y} + \epsilon \frac{q^2}{m_{0} \Phi^3}\frac{\partial \Phi}{\partial y}.
\end{equation}
Also,
\begin{equation}
\label{m dot}
\frac{dm_{0}}{ds} =  \epsilon \frac{q^2}{m_{0}\Phi^3}u^{\mu}\frac{\partial \Phi}{\partial x^{\mu}} - \epsilon \frac{q}{2 \Phi^2}u^{\mu}u^{\nu}\frac{\partial g_{\mu \nu}}{\partial y} - \epsilon \frac{q^2}{m_{0} \Phi^2}u^{\mu}\left(\frac{\partial A_{\mu}}{\partial y} + \frac{A_{\mu}}{\Phi}\frac{\partial \Phi}{\partial y}\right).
\end{equation}
It is important  to mention that the above expressions, (\ref{q dot}) and (\ref{m dot}), are invariant under the set of ``gauge" transformations
\begin{eqnarray}
\label{Allowed transformations}
x^{\mu} &=& \bar{x}^{\mu},\nonumber \\ y^{4}&=& \bar{y}^{4}+ f(\bar{x}^{0},\bar{x}^{1},\bar{x}^{2},\bar{x}^{3}),
\end{eqnarray}
that keep the shape of the line element (\ref{special metric}) invariant. 

Let us immediately  notice that, in the case of ``classical" Kaluza-Klein theory, we recover some well know results. If the    extra dimension is compactified, that is, if there is  no dependence on the extra coordinate, then from (\ref{q dot}) it follows that $q$ (or $P_{4}$) is strictly constant. On the other hand (\ref{m dot}) can be easily integrated to recover  (\ref{rel for m, M and q}). Which implies that the  rest mass varies as $\Phi^{-1}$. In particular, if there is no scalar field, then $m_{0}$ is also a constant of motion.  

In any other situation where the extra dimension is large, or non-compactified, it follows from (\ref{q dot}) that $q = P_{4}$ will change along the trajectory of the particle. This is a  totally new feature. It was  missed in classical versions of Kaluza-Klein theory because of the imposition of the so-called cylinder condition. When this condition is dropped,  both parameters, $q$ and $m_{0}$, will change along the trajectory. 

\medskip

If the quantity $q$ can be identified with the electric charge $q_{e}$, then the electric charge and rest mass are not conserved separately. The conserved  quantity along the motion of the particle, is now the combination $ m_{0}^2 + \epsilon q_{e}^2/\Phi^2$.

\medskip 

 Second, we consider massless test particles moving in $5D$. For $\epsilon = +1$, such particles are observed in $4D$ as the motion of massless particles with $q = 0$.   If the extra dimension is spacelike, then it follows from (\ref{massless 5D particles}) that massless $5D$-particles are observed in $4D$ as having effective rest mass $m_{0} = |q|/\Phi$. If $q = 0$, the motion observed in $4D$ is along isotropic geodesics ($ds = 0$) in spacetime. 

At this point one could ask,  how can that be? If $M_{(5)} = 0$, where does the observed effective rest mass come from? In answering this question the identification of $q$ with the electric charge is helpful. With this interpretation, the observer in $4D$ perceives the  effective rest mass $m_{0}$ as being  totally of electromagnetic origin, in the  usual classical sense that setting $q_{e} = 0$ requires  $m_{0} = 0$ \cite{TRK}-\cite{EMM2}. 

\subsection{Variation of $\alpha_{em}$ and $\sigma$}

If we assume that the speed of light $c$ and $\hbar$ are constants, then the variation of the fine-structure constant $\alpha_{em}$ is a consequence of the variation of electric charge. With the interpretation $q = q_{e}$, the change  of $\alpha_{em}$, in explicit form, can be written as   

\begin{equation}
\label{fine structure constant, explicit expression}
\frac{1}{\alpha_{em}}\frac{d\alpha_{em}}{ds} = \epsilon \left(\frac{dy}{ds} + A_{\lambda}u^{\lambda}\right)^{- 1}\Phi^{-2}u^{\mu}u^{\nu}\frac{\partial g_{\mu \nu}}{\partial y} + 2 u^{\mu}
\frac{\partial A_{\mu}}{\partial y} + 2 \left(\frac{dy}{ds} + A_{\lambda}u^{\lambda}\right)\frac{1}{\Phi}\frac{\partial \Phi}{\partial y},
\end{equation}
evaluated along the trajectory of the particle.  Notice that only derivatives with respect to the extra variable appear in this equation, there are no derivatives with respect to the spacetime coordinates. Thus, a large extra dimension leads naturally to a variation of the fine structure constant. 

For the Thomson cross section $\sigma$ we have
\begin{equation}
\label{Thomson cross section}
\frac{1}{2 \sigma} \frac{d \sigma}{ds} = \frac{1}{\alpha_{em}}\frac{d\alpha_{em}}{ds} - \frac{1}{m_{0}}\frac{dm_{0}}{ds}.
\end{equation} 
There is no need to write its explicit expression. It is clear that $\sigma$ changes even in the case of compactified extra dimension.

\subsection{Relationship between the variation of $q$ and $m_{0}$}

Equations (\ref{q dot}) and (\ref{m dot}) are quite complicated and it is difficult to see from them if there is any connection between $dq/ds$ and $dm_{0}/ds$. With this in mind, let us consider the class of simplified metrics 

\begin{equation}
\label{simplified metric}
d{\cal{S}}^2 = g_{\mu\nu}(x^{\rho}, y)dx^{\mu}dx^{\nu} + \epsilon dy^2,
\end{equation}
which are popular in brane-world as well as in space-time-matter theories. They allow us  to isolate the effects of the large extra dimension, without the details induced by the electromagnetic and scalar field.  For these metrics, (\ref {q dot}) and (\ref{m dot}) reduce to 
\begin{equation}
\label{q dot simplified}
\frac{dq}{ds} = \frac{1}{2}m_{0}u^{\mu}u^{\nu}\frac{\partial g_{\mu\nu}}{\partial y},
\end{equation}
and 
\begin{equation}
\label{m dot simplified}
\frac{dm_{0}}{ds} =  - \epsilon \frac{1}{2} q u^{\mu}u^{\nu}\frac{\partial g_{\mu \nu}}{\partial y}. 
\end{equation}
Using (\ref{u4}) and (\ref{q in terms of m and Phi}), this equation can be written as
\begin{equation}
\label{dm/m}
\frac{1}{m_{0}}\frac{dm_{0}}{ds} = - \frac{1}{2}u^{\mu}u^{\nu}\frac{\partial g_{\mu\nu}}{\partial y} \frac{dy}{ds}.
\end{equation}

The effects of a large extra dimension on $q = P_{4}$ and $m_{0}$ are closely related to the extra (or fifth) force on test particles. They are proportional to the first derivatives of the metric with respect to the extra, non-compactified coordinate \cite{Youm1}-\cite{Seahra3}. Indeed, although there is some controversy in the literature, this extra force is given by  
\begin{equation}
\label{fifth force}
f^{\sigma} = [u^{\sigma}u^{\lambda} - g^{\sigma \lambda}]u^{\rho}\frac{\partial g_{\lambda \rho}}{\partial y} \frac{dy}{ds}.
\end{equation}
On the other hand, the expression (\ref{dm/m}) is {\em identical} to the one obtained, from different considerations \cite{Youm1}, in the discussion of a component of the fifth force that is parallel to the four-velocity of the test particle (in that discussion (\ref{fifth force}) would be the perpendicular component). 

Let us now introduce the parameter $w$ as 
\begin{equation}
\label{w}
w = \frac{1}{2}\int{\left(\frac{\partial g_{\mu\nu}}{\partial y}u^{\mu}u^{\nu}\right) ds}.
\end{equation}
Equations (\ref{q dot simplified}) and (\ref{m dot simplified}) become
\begin{equation}
\frac{dq}{dw} = m_{0},
\end{equation}
and
\begin{equation}
\frac{dm_{0}}{dw} = - \epsilon q.
\end{equation}
With this parameterization, $q$ and $m_{0}$ behave as ``conjugate coordinates", satisfying the equations
\begin{equation}
\label{armonic motion for q}
\frac{d^{2} q}{dw^2} + \epsilon q = 0,
\end{equation}
and $d^{2}m_{0}/dw^2 + \epsilon m_{0} = 0$. These equations, with {\em positive} $\epsilon$, remind of the mechanical harmonic motion. Within this analogy,  $q$ would play the role of the displacement from equilibrium ($q = 0$), $m_{0}$ the velocity, and $d^{2}q/dw^2 = - \epsilon q$ the ``restoring" force. The first integral of (\ref{armonic motion for q}) would give back the conservation equation (\ref{rel for m, M and q}), i.e.,  $(dq/dw)^2 = M_{(5)}^2 - q^2$. Thus, $q$ would ``oscillate" with amplitude $M_{(5)}$, viz.,
\begin{equation}
\label{solution for q}
q = M_{(5)}\sin(w - \bar{w}), 
\end{equation}
where $\bar{w}$  is a constant of integration. Similarly, for the rest mass 
\begin{equation}
\label{solution for m}
m_{0} = M_{(5)}\cos(w - \bar{w}). 
\end{equation}
Now, for a spacelike ($\epsilon = -1$) large extra dimension, the mechanical analog would correspond to an elliptical motion
\begin{eqnarray}
\label{q and m for epsilon - 1}
m_{0} &=& M_{(5)}\cosh[w - \bar{w}],\nonumber \\
q &=&  M_{(5)}\sinh[w - \bar{w}]. 
\end{eqnarray}

The above analogy can be formally extended to test particles moving in an arbitrary $5D$ bulk metric, with non-vanishing electromagnetic potentials and $\Phi = 1$. Indeed, from  (\ref{rel for m, M and q}) it follows that one can formally set  
\begin{eqnarray}
\label{m and q general}
m_{0} &=& {\tilde{m}}_{0}\cos(\tilde{w} - {\tilde{w}}_{0}) - \tilde{q}\sin(\tilde{w} - {\tilde{w}}_{0}),\nonumber \\
q &=& {\tilde{m}}_{0}\sin(\tilde{w} - {\tilde{w}}_{0}) + \tilde{q}\cos(\tilde{w} - {\tilde{w}}_{0}),
\end{eqnarray}
for $\epsilon = + 1$, and 
\begin{eqnarray}
m_{0} &=& {\tilde{m}}_{0}\cosh(\tilde{w} - {\tilde{w}}_{0}) + \tilde{q}\sinh(\tilde{w} - {\tilde{w}}_{0}),\nonumber \\
q &=& {\tilde{m}}_{0}\sinh(\tilde{w} - {\tilde{w}}_{0}) + \tilde{q}\cosh(\tilde{w} - {\tilde{w}}_{0}),
\end{eqnarray}
for $\epsilon = -1$, provided the parameter $\tilde{w}$ is a solution of the equation
\begin{equation}
\frac{d \tilde{w}}{ds} = \frac{\partial g_{\mu\nu}}{\partial y}u^{\mu}u^{\nu} + \frac{q}{m_{0}}u^{\mu}\frac{\partial A_{\mu}}{\partial y}.
\end{equation}
The constants ${\tilde{m}}_{0} = m_{0}({\tilde{w}}_{0})$ and $\tilde{q} = q({\tilde{w}}_{0})$ satisfy ${\tilde{m}}_{0}^2 + \epsilon {\tilde{q}}^2 = M_{(5)}^2$. It is clear that the above simple mechanical analog is a unique consequence of  the dependence of five-dimensional metrics on the extra coordinate. This is the theoretical framework in brane-world theories and STM.

At this point we notice that in a recent paper Wesson \cite{2times} discusses five-dimensional general relativity with two ``times", i.e., in spacetimes with an extra  timelike dimension. For null geodesics in $5D$, using ``canonical" coordinates he shows that the parameter $l$, which is related to the rest mass of test particles in $4D$, oscillates in a simple harmonic motion, similar to our equations (\ref{solution for m}) and (\ref{m and q general}). Therefore, our approach can be used to extend Wesson's discussion to massive particles in $5D$, with the advantage that here we employ an invariant definition for the rest mass $m_{0}$. 

We would like to finish this section with the following comment. Since the last term in (\ref{q dot}) is proportional to $\partial \Phi/ \partial y$, the question may arise of whether the consideration of the metric
\begin{equation}
\label{simplified metric with Phi(y)}
d{\cal{S}}^2 = g_{\mu\nu}(x^{\rho})dx^{\mu}dx^{\nu} + \epsilon \Phi^2(y)dy^2,
\end{equation}
would not induce a spurious variation of $q$. This question is similar to ask whether the energy would be conserved along the motion of a test particle in a spacetime with metric $ds^2 = g_{00}(t)dt^2 + g_{ij}(x^1, x^2, x^3)dx^i dx^j$. The energy equation would be $dP_{0}/ds = (m_{0}/2)(\partial g_{00}/\partial t)(dt/ds)^2$. In this case, it is clear that the ``no-conservation" of energy is not a result of any physical mechanism, but it is due to the bad choice of time coordinate. The same discussion can be applied for a charged test particle moving in the background metric (\ref{simplified metric with Phi(y)}). For this metric,  Eq. (\ref{q dot}) is equivalent to $dP_{4}/d{\cal {S}} = dq/d{\cal {S}} = (M_{(5)}/2)(\partial \gamma_{44}/\partial y) (dy/d{\cal {S}})^2$. Once again, the ``wrong" choice of coordinate $y$ leads to a spurious change of charge that can be eliminated by choosing a new coordinate $d{\bar{y}} = \Phi(y)dy$. 

\section{Variation of charge and rest mass in a cosmological setting} 

The goal of this section is twofold. Firstly, to present a simple example that illustrates the usefulness of the Hamilton-Jacobi method for the invariant definition of rest mass. Secondly, to examine the variation of charge, rest mass, $\alpha_{em}$ and $\sigma$ in a cosmological setting. 

We will consider the motion of test particles in the background metric \cite{JPdeL 1} 
\begin{equation}
\label{Ponce de Leon solution}
d{\cal S}^2 = y^2 dt^2 - t^{2/\alpha}y^{2/(1 - \alpha)}[dr^2 + r^2(d\theta^2 + \sin^2\theta d\phi^2)] - \alpha^2(1- \alpha)^{-2} t^2 dy^2,
\end{equation}
where $\alpha$ is a constant, $y$ is the coordinate along the  extra-dimension and $t, r, \theta$ and $\phi$ are the usual coordinates for a spacetime with spherically symmetric spatial sections. This is a solution to the five-dimensional Einstein field equations, with ${^{(5)}T}_{AB} = 0$. 

In four-dimensions (on the hypersurfaces $y = const.$) this metric corresponds to the $4D$ Friedmann-Robertson-Walker models with flat $3D$ sections. The  energy density $\rho_{eff}$  and pressure $p_{eff}$ of the effective $4D$ matter satisfy the equation of state

\begin{equation} 
\label{eq of state for the eff fluid}
p_{eff} = n \rho_{eff},  
\end{equation}
where $n = ({2\alpha}/{3} -1)$. Thus  for $\alpha = 2$ we recover radiation, for $\alpha = 3/2$  we recover dust, etc.

 In spherically symmetric fields test particles move on  a single ``plane" passing through the center. We take this plane  as the $\theta = \pi/2$ plane. Then, the Hamilton-Jacobi equation, for the metric (\ref{Ponce de Leon solution}) is 
\begin{equation}
\frac{1}{y^2}\left(\frac{\partial S}{\partial t}\right)^2 - \frac{1}{t^{2/\alpha}y^{2/(1 - \alpha)}}\left[\left(\frac{\partial S}{\partial r}\right)^2 + \frac{1}{r^2}\left(\frac{\partial S}{\partial \phi}\right)^2\right] - \frac{(1 - \alpha)^2}{\alpha^2 t^2}\left(\frac{\partial S}{\partial y}\right)^2 = M_{(5)}^2.
\end{equation}
It is clear that the action separates as
\begin{equation}
S = S_{1}(t, y) + S_{2}(r) + L \phi,
\end{equation}
where $L$ is the angular momentum.  
Thus, we obtain
\begin{equation}
\label{S1}
\frac{1}{y^2}\left(\frac{\partial S_{1}}{\partial t}\right)^2 - \frac{k^2}{t^{2/\alpha}y^{2/(1 - \alpha)}}  - \frac{(1 - \alpha)^2}{\alpha^2 t^2}\left(\frac{\partial S_{1}}{\partial y}\right)^2 = M_{(5)}^2,
\end{equation}
and
\begin{equation}
\left(\frac{dS_{2}}{dr}\right)^2 + \frac{L^2}{r^2} = k^2 \geq 0,
\end{equation}
where $k$ is the separation constant. If it is zero, then the particle is commoving (or at rest) in the system of reference defined by (\ref{Ponce de Leon solution}).

In this case, the spacetime basis vectors (\ref{associated basis vectors}) reduce to ${\hat{e}}_{(\mu)}^A = \delta_{\mu}^A$. Thus, from (\ref{four momentum}) and (\ref{def of rest mass}) we obtain the effective rest mass, as measured in $4D$, as follows
\begin{equation}
\label{effective mass as measured in 4D}
m_{0}^2 = \frac{1}{y^2}\left(\frac{\partial S_{1}}{\partial t}\right)^2 - \frac{k^2}{t^{2/\alpha}y^{2/(1 - \alpha)}},
\end{equation}
evaluated at the trajectory $y = y(t)$. We stress the fact that this expression is totally free of the ambiguities, induced by the choice of parameters used to characterize the motion in $5D$ and $4D$, typical of the approach where the equations of motion for test particles in $4D$ are obtained from the geodesic equation in $5D$. 

The relative variation of $m_{0}$, for particles fixed in space $u^{\mu} = {\delta}^{\mu}_{0}/y$, is given by
\begin{equation}
\label{rel. variation of rest mass in cosmological setting}
\frac{1}{m_{0}}\frac{dm_{0}}{ds} = - \frac{1}{y}\frac{dy}{ds}\left[1 + \frac{\alpha^2 t}{(1 - \alpha)^2}\frac{dy}{ds}\right].
\end{equation}
Also, for the fine structure constant we get
\begin{equation}
\label{variation of the fine structure constant in cosmological setting}
\frac{1}{\alpha_{em}}\frac{d\alpha_{em}}{ds} = - \frac{2(1 - \alpha^2)^2}{\alpha^2 t^2 y}\left(\frac{dy}{ds}\right)^{- 1}.
\end{equation} 
These quantities have to be evaluated along the particle's worldline. The variation of $\sigma$ is then obtained from (\ref{Thomson cross section}).

 At this point, it is worth noticing that in the case where $q = P_{4} = - (\partial S/\partial y) = 0$, the equation of motion (\ref{S1}) requires  $y = const$, and the effective rest mass observed in $4D$ is  $m_{0} = M_{(5)}$. In other words a particle moving in $5D$ with no momentum in the extra direction, which is observed in $4D$ as the motion of a particle with no-electric charge, will be insensitive to the extra dimension.  Interesting results appear when the $5D$ motion has nonzero $P_{4}$-component.  

Let us now consider different physical scenarios allowed in (\ref{S1}) 

\subsection{Massive test particles: $M_{(5)} \neq 0$}
In this case there are two possibilities. Either the particle is at rest ($k = 0$), or it is somehow moving in spacetime ($k \neq 0$). In order to isolate the effects from the extra dimension, from the effects due to the motion in spacetime, we will consider $k = 0$.

From (\ref{S1}) one can easily get
\begin{equation}
S_{1}(t, y) = \pm \frac{\alpha M_{(5)}}{\sqrt{2\alpha - 1}} yt.
\end{equation}
Consequently, 
\begin{equation}
m_{0} = \frac{\alpha M_{(5)}}{\sqrt{2\alpha - 1}}, \;\;\;\;\;q =  \pm m_{0}t.
\end{equation}
Here the rest mass is constant, because of the mutual cancelation of the change induced by the term $(\partial g_{\mu\nu}/\partial y)u^{\mu}u^{\nu}$ and the change induced by the scalar field. 

Now, using 
\begin{equation}
\label{Momentum}
P^A = M_{(5)}\frac{d\xi^A}{d{\cal{S}}} = \gamma^{AB}P_{B} = - \gamma^{AB}\frac{\partial S}{\partial \xi^B},
\end{equation}
we obtain
\begin{equation}
\frac{dt}{d\cal{S}}  = \mp \frac{\alpha}{y \sqrt{2\alpha -1}},
\end{equation}
and
\begin{equation}
\frac{dy}{d\cal{S}} = \pm \frac{(1 - \alpha)}{t\sqrt{2\alpha -1}}.
\end{equation}
From these expressions we evaluate $dy/dt$ and integrate to obtain
\begin{equation}
\label{trajectory}
y = Dt^{(\alpha - 1)/\alpha},
\end{equation}
along the trajectory. Here $D$ is a constant of integration. The proper time $\tau$ along the trajectory is given by $d\tau = y(t)dt$. Thus, from (\ref{trajectory}), we get $\tau \sim t^{(2 \alpha -1)/\alpha}$. Consequently, in this case, $m_{0}$ is a constant and $q \sim \tau^{\alpha/(2\alpha -1)}$. 

Since $m_{0}$ is constant here, it follows from (\ref{Thomson cross section}) that the change of the Thomson cross section $\sigma$ is similar to that of  $\alpha_{em}$, which is given by 

\begin{equation}
\label{fine structure constant for massive 5D particles}
\frac{1}{\alpha_{em}}\frac{d\alpha_{em}}{d\tau} = \frac{2 \alpha^2}{(2\alpha - 1)\tau}.
\end{equation}
Models with $\alpha < 1$ have inflationary equations of state. During the cosmological ``constant" era evolution $p_{eff} \approx - \rho_{eff}$  ($\alpha \rightarrow 0$), $\alpha_{em}$ and $\sigma$ decrease with the increase of $\tau$. Then, in the curvature era, and further in the radiation and dust dominated era  $\alpha_{em}$ and $\sigma $ grow as a power of $\tau$. 

\subsection{Massless test particles: $M_{(5)} = 0$}

In this case the trajectory in $5D$ is along isotropic geodesics. These are given by the Eikonal equation, which is formally obtained from the above formulae by setting $M_{(5)} = 0$ in (\ref{S1}). According to the discussion in section 2.2.2, there are two different physical possibilities here. They are   $q \neq 0$, or $q = 0$. 

\subsubsection{$M_{(5)} = 0$, $q \neq 0$}

For the same reason as above, we will consider $k = 0$. Then, equation (\ref{S1}) separates and, we obtain
\begin{equation}
S_{1} = C t^{\pm l}y^{\pm l \alpha/(1 - \alpha)},
\end{equation} 
where $C$ is a constant of integration and $l$ is the separation constant. Now in (\ref{Momentum}) instead of the derivatives $M_{(5)}d/d{\cal{S}}$ we have to write derivatives $d/d\lambda$, where $\lambda$ is the parameter along the null geodesic. Thus, we find
\begin{equation}
\frac{dt}{d\lambda} = \mp C l t^{(- 1 \pm l)}y^{(- 2 \pm l \alpha/(1 - \alpha))},
\end{equation}
and
\begin{equation}
\frac{dy}{d\lambda} = \pm C l \frac{1 - \alpha}{\alpha}t^{(- 2 \pm l)} y ^{(- 1 \pm l \alpha/(1 - \alpha))}.
\end{equation}
From the above expressions we evaluate $dy/dt$ and integrate to obtain $y = Dt^{(\alpha - 1)/\alpha}$. This is the same equation as in the massive case (\ref{trajectory}). However, in this case we find $q$ and $m_{0}$ as follows
\begin{eqnarray}
q &=& \mp A \alpha (1 - \alpha)^{- 1}t^{(1 - \alpha)/\alpha},\nonumber \\
m_{0} &=& A t^{(1 - 2\alpha)/\alpha},
\end{eqnarray}
where $A$ is expressed through the other constants as $A = ClD^{(- 1 \pm l \alpha /(1 - \alpha))}$. Thus, here $q \sim \tau^{(1 - \alpha)/(2\alpha - 1)}$ and $m_{0} \sim \tau^{- 1}$. Thus,
\begin{equation}
\label{Thomson cross section anf fine structure constant for massless 5D particles}
\frac{1}{\sigma}\frac{d\sigma}{d\tau} = \frac{2}{(2\alpha - 1)\tau}, \;\;\;\; \frac{1}{\alpha_{em}}\frac{d\alpha_{em}}{d\tau} = \frac{2(1 - \alpha)}{(2\alpha - 1)\tau}.
\end{equation}
Near the vacuum dominated era $p_{eff} \approx - \rho_{eff}$, $\alpha_{em}$ and $\sigma$ decrease with the increase of  $\tau$. In the quintessential epoch for which $-2\rho_{eff}/3 < p_{eff} < - \rho_{eff}/3$ ($1/2 < \alpha < 1$), $\alpha_{em}$  and $\sigma$ grow with time. In the very early universe ($\alpha = 3$), radiation epoch ($\alpha = 2$) and dust era ($\alpha = 3/2$), $\alpha_{em}$ decreases and $\sigma$ increases as a power of the proper time $\tau$. 

Hence there are very different possibilities for the change in $\alpha_{em}$ or $q_{e}$ depending on whether the test particles in $5D$ are assumed massive ($M_{(5)} \neq 0$) or massless ($M_{(5)} = 0$). However, the Thomson cross section $\sigma$ has a similar  behavior in both cases, except in the quintessential epoch.  

\subsubsection{$M_{(5)} = 0$, $q = 0$}

In this case the motion, as observed in $4D$, is lightlike and, therefore, $k$ must be different
from zero. The left hand side in (\ref{effective mass as measured in 4D})  is $m_{0} = 0$. This equation is equivalent to $k^{\mu}k_{\mu} = 0$, where $k_{\mu}$ is the $4D$ wave vector.  Since $P_{4} = 0$, it follows that $y = const$ along the motion.  Therefore, the frequency $\omega = - \partial S/ \partial t$ of the ``induced" photon is 

\begin{equation}
\omega \sim \tau^{- 1/\alpha}.
\end{equation} 

For completeness, we mention that there is one more possibility left here. Namely the motion in $5D$ with $d{\cal{S}}^2 < 0$. In this case replacing $M_{(5)}^2 \rightarrow - {\bar{M}}^2$ we obtain the same expressions as in section 4.1, but with $\bar{M}$ instead of $M_{(5)}$ and $\sqrt{1 - 2\alpha}$ instead of $\sqrt{2\alpha - 1}$. Thus,  such a five-dimensional motion is interpreted by an observer in $4D$ as a test particle with positive effective rest mass. This is a pure consequence of the motion in $5D$ along the {\em spacelike} extra coordinate.  

\section{Summary and Conclusions}

We have given the general formula for the change of the rest mass of a test particle, as measured by an observer in $4D$. Equation (\ref{m dot}) generalizes to an arbitrary five-dimensional metric the particular  expression (\ref{dm/m}) for the metric in gaussian normal coordinates (\ref{simplified metric}).
The variation of rest mass is predicted  in classical versions of compactified Kaluza-Klein theory with scalar field. The last three terms in (\ref{m dot}) are the ``corrections" induced by a large extra dimension. 

However, the electric charge is strictly constant in compactified versions of Kaluza-Klein theory. This is because the cylinder condition requires the extra coordinate $y$ to be cyclic. Then, according to the Hamilton-Jacobi formalism,  the action separates as $S = S(x^{\mu}) - q y$, where $q = P_{4} = - \partial S/\partial y = const$. When this condition is dropped, $q = P_{4}$ will vary along the trajectory according to (\ref{q dot}). Thus, if $q$ is identified with the electric charge, then an observer in $4D$ will find ``inconsistencies" with the law of conservation of charge in $4D$. In the present scenario the conserved quantity along the trajectory is the combination $m_{0}^2 + \epsilon q_{e}^2/\Phi^2$.

The variation of $m_{0}$ and $q_{e}$ has important implications in a wide variety of phenomena. In particular, in the scattering of electromagnetic radiation by charged particles, for which the cross section is proportional to $q_{e}^4/m_{0}^2$. Also in the variation of the fine structure constant $\alpha_{em}$, recently observed in the study of quasar absorption line spectra.

The discussion in section 4 shows  how to obtain the rest mass and $q$ (or charge), as observed in $4D$, from the Hamilton-Jacobi formalism in $5D$. It also  illustrates the different physical scenarios perceived by an observer in $4D$. We have seen that the effective $4D$ quantities depend not only on the motion in $5D$ but also on whether the $5D$ test particle is massive or massless. The results obtained for $m_{0}$, $q_{e}$, $\alpha_{em}$ and $\sigma$ are intended to be illustrative rather than experimental suggestions.  

The question may arise of whether the predicted variation of rest mass and charge contradicts some well established assumptions in physics, like charge conservation. In our interpretation, the answer to this question is negative. Indeed, since the variation of these quantities takes place on cosmic timescales, they would not be observed directly in laboratory. This seems to be a general feature of constants in higher dimensional theories whose effective value in $4D$ may vary in time and space. This is similar to what happens in cosmological models that are free of singularities in $5D$, but singular when interpreted in $4D$ \cite{Inevof sing}. 

Our analysis also clarifies the $4D$ interpretation of null geodesics in $5D$. We have seen that such geodesics appear as timelike paths in $4D$ only if the following two conditions are met {\em simultaneously}: (i) the extra dimension is spacelike, and (ii) the particle in its five-dimensional motion has $P_{4} \neq 0$ (or $q \neq 0$). Otherwise, a null geodesic in $5D$ is observed as a lightlike particle in $4D$. 

We notice that in our discussion the underlying physics motivating the introduction of  a large extra dimension was nowhere used. Neither, the physical meaning of the extra coordinate. Therefore, our  results are applicable to brane-world models, STM, and other $5D$ theories with a large extra dimension\footnote{In STM and in the thick brane scenario the $5D$ manifold is smooth everywhere and there are no defects. In the RS2 brane-world scenario \cite{RS2} our universe is a singular $4D$ hypersurface and the derivatives $\partial g_{\mu \nu}/\partial y$ are discontinuous, and change sign, through the brane. However, the discontinuity is not observed \cite{Seahra3} and effective $4D$ equations can be obtained by taking mean values and applying Israel's junction conditions through the brane \cite{Youm2}. }.

Thus, the possibility that our $4D$ universe is embedded in a higher dimensional bulk space, with more than four non-compact extra dimensions, should have major consequences for astrophysics and cosmology. The existence of an extra (fifth) force is one of them. Here we presented the variation of the fundamental electron charge. Elsewhere we discussed the variation of  the  gravitational coupling $G$ and the cosmological term $\Lambda_{(4)}$ \cite{VarG}.
In principle, these effects are observable, but theoretical predictions are model-dependent. Therefore, other $5D$ metrics have to be studied and tested experimentally for compatibility with observational data.

\end{document}